\documentclass[aip,jmp,showpacs,amsmath,amssymb,floatfix,10pt]{revtex4-1} 

\usepackage[pdftex]{graphicx}       
\usepackage{txfonts}        
\usepackage{amsmath}
\usepackage{amssymb}
\usepackage{mathtools}
\usepackage{mathrsfs}	

\newcommand{\Ai}{\operatorname{Ai}}

\usepackage{hyperref} 
\hypersetup{colorlinks,citecolor=blue,filecolor=blue,linkcolor=blue,urlcolor=blue}

\begin{document}
\title{Exact solution of pulled, directed vesicles with sticky walls in two dimensions}
\author{A. L. Owczarek}\email{owczarek@unimelb.edu.au}
\affiliation{School of Mathematics and Statistics, University of Melbourne, Victoria 3010, Australia}
\author{T. Prellberg} \email{t.prellberg@qmul.ac.uk}
\affiliation{School of Mathematical Sciences, Queen Mary University of
  London, Mile End Road, London, E1 4NS, United Kingdom}
\date{\today}

\begin{abstract}

We analyse a directed lattice vesicle model incorporating both the binding-unbinding transition and the vesicle inflation-deflation transition. From the exact solution we derive the phase diagram for this model and elucidate scaling properties around the binding-unbinding critical point in this larger parameter space. We also consider how the phase diagram changes when a perpendicular force is applied to the end of a directed vesicle. 

\end{abstract}
\maketitle

\section{Introduction}
The modelling of biological vesicles, where there can be an internal osmotic pressure, using various different lattice walks and polygons on two-dimensional lattices has proved to demonstrate mathematically rich and physically faithful results \cite{leibler1987a-a,brak1990a-a,fisher1991a-a,owczarek1993a-:a,brak1994a-:a,prellberg1995d-a,brak1998c-:a,prellberg1999a-:a,janse2000a-a,cardy2001a-a,richard2001a-a,richard2002a-a,janse2003a-a,cardy2003a-a,richard2004a-a,richard2006a-a}.  Self-avoiding polygons enumerated by perimeter and area \cite{leibler1987a-a,fisher1991a-a,cardy2001a-a,richard2001a-a,richard2002a-a,cardy2003a-a,richard2004a-a} provides the fundamental model in this field. The pressure conjugate to the area drives an inflation-deflation transition in the vesicle. However, results for this model need to be calculated approximately so other simpler models have been used to calculate exact results. An exactly solvable model that has received most recent attention mathematically \cite{owczarek2012a-:a,owczarek2012b-:a,owczarek2014a-:a,haug2015a-a} is that of Dyck paths above a wall with various extensions to consider the competition with pulling forces.  A more appealing model physically is that of so-called staircase polygons enumerated in the same way \cite{brak1990a-a,bousquet1992a-a,prellberg1995d-a,richard2006a-a}. Without  enumeration by area, and so ignoring the contribution of pressure, various exact models of two directed paths, which also form configurations of staircase polygons have been recently considered \cite{owczarek2012c-:a,tabbara2014a-:a,owczarek2017a-:a}. In some cases \cite{tabbara2014a-:a,owczarek2017a-:a} an attractive force between the two walks was considered. An attractive force can bind the two walks together at low temperatures and so if applied to a vesicle model with pressure can potentially compete with the inflation-deflation transition. The expansion of the model from strictly staircase polygons to two directed walks that start and end together to form a vesicle but can touch along the way allows for the consideration of a model with both pressure and an attractive contact interaction between the sides of the vesicle. Indeed, that is precisely what we consider in this work. For completeness we have also considered a model where one end of the vesicle is pulled perpendicular to the direction of the main diagonal of the vesicle. We solve this model exactly and analyse the solution to elucidate the effects of pulling and binding on the inflation-deflation transition.

\section{The model}

\begin{figure}[ht]
\includegraphics[height=6cm]{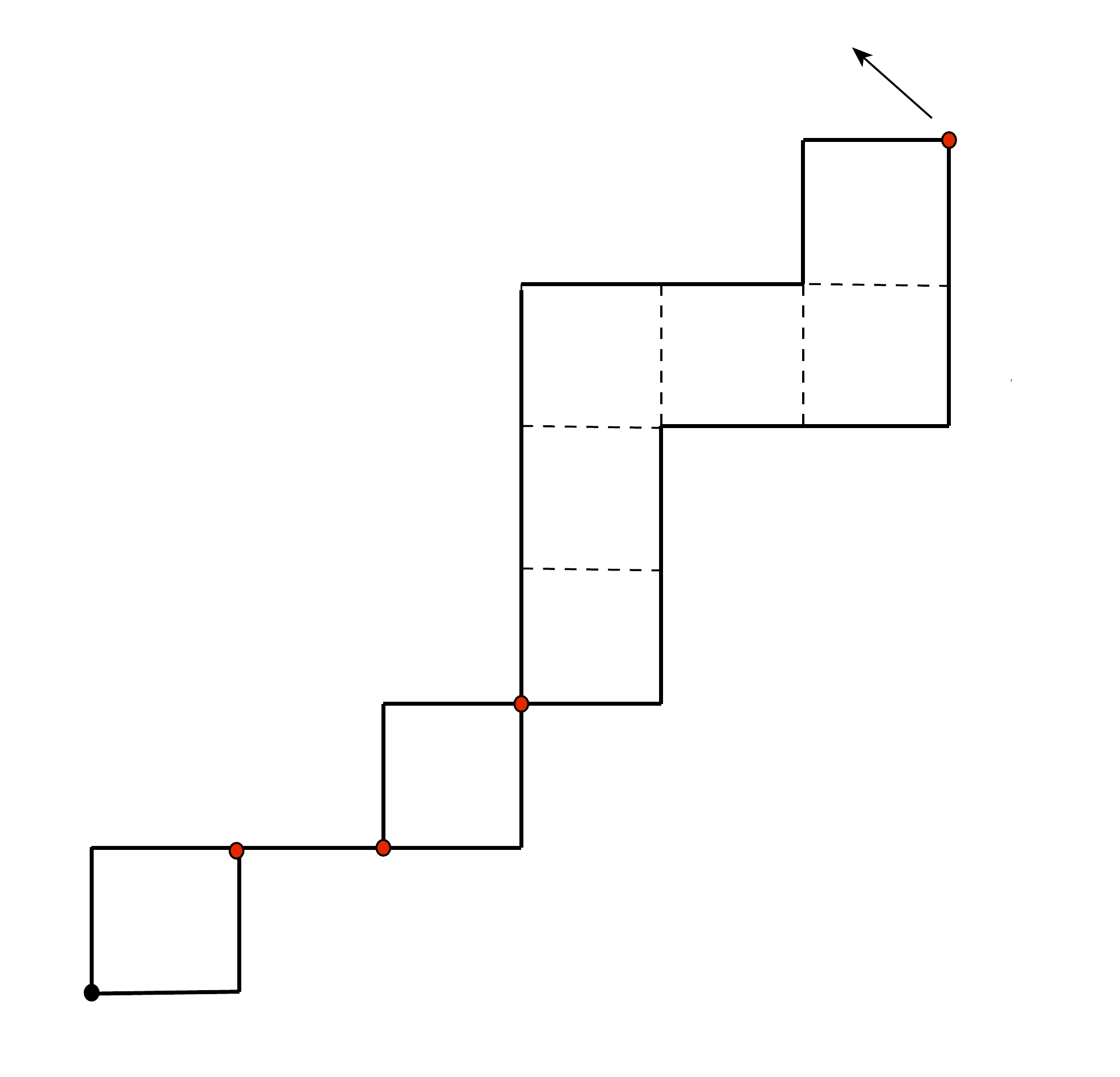}
\caption{Two directed walks starting and ending together each of length $n=12$. There are $a=8$ plaquettes and $m_c=4$ shared/common sites excluding the origin.  An arrow indicates the direction of the pulling force we consider.}
\label{fig1}
\end{figure}

We consider a pair of non-crossing lattice paths on $\mathbb Z^2$ taking steps only in the east $(1,0)$ or north $(0,1)$ directions, starting at the origin $(0,0)$ and ending at a common point $(n_x,n_y)$. Such walks are typically referred to as (infinitely) friendly walks. We weight these pairs of walks by the number $a$ of enclosed plaquettes and the number $m_c$ of common sites excluding the origin, as well as the deviation $h=n_y-n_x$ of the end point from the main diagonal. Let $\Omega$ denote the class of all such pairs of walks of arbitrary length. An example is given in Figure \ref{fig1}.

We define the partition function
\begin{equation}
Z_n(c,s,q)=\sum_{\varphi\in\Omega,\;|\varphi|=n}c^{m_c(\varphi)}s^{h(\varphi)}q^{a(\varphi)}\;,
\end{equation}
where $|\varphi|$ denotes the length of the configuration $\varphi$, which is given by the end point position via $n_x+n_y$. We form the generating function
\begin{equation}
G(c,s,q,t)=\sum_{n=0}^\infty t^nZ_n(c,s,q)\;.
\end{equation}
The reduced free energy is given by
\begin{equation}
\psi(c,s,q)=-\lim_{n\to\infty}\frac1n\log Z_n(c,s,q)
\end{equation}
and is related to the smallest positive singularity $t_c(c,s,q)$  of the generating function $G(c,s,q,t)$ via
$\psi(c,s,q)=\log t_c(c,s,q)$.

Configurations that don't share any edges are known as staircase polygons \cite{brak1990a-a}. The generating function $S(x,y,q)$ for staircase polygons enumerated by height, width and the number of enclosed plaquettes is then given by
\begin{equation}
S(x,y,q)=\sum_{\varphi\in\Omega,\;m_c(\varphi)=0}x^{n_x(\varphi)}y^{n_y(\varphi)}q^{a(\varphi)}\;,
\end{equation}
For the derivation of the exact solution it is more natural to work in the variables $x$ and $y$ instead of $s$ and $t$, and we therefore define 
\begin{equation}
F(c,x,y,q)=\sum_{\varphi\in\Omega}c^{m_c(\varphi)}x^{n_x(\varphi)}y^{n_y(\varphi)}q^{a(\varphi)}\;,
\end{equation}
so that
\begin{equation}
\label{Gsubs}
G(c,s,q,t)=F(c,t/s,ts,q)\;.
\end{equation}

\section{Exact Solution}

Any configuration in $\Omega$ can be written as a sequence of translated staircase polygons and pairs of horizontal and vertical steps. A standard necklace argument therefore gives
\begin{equation}
F(c,x,y,q)=\frac1{1-c[x+y+S(x,y,q)]}\;.
\end{equation}
The generating function for staircase polygons is well known \cite{brak1990a-a}. A standard combinatorial decomposition leads to the functional equation
\begin{equation}
\label{functeq}
S(x,y,q)=[qx+S(qx,y,q)][y+S(x,y,q)]\;,
\end{equation}
from which one can obtain the explicit solution as
\begin{equation}
S(x,y,q)=y\left[\frac{H(qx,y,q)}{H(x,y,q)}-1\right]\;,
\end{equation}
where
\begin{equation}
H(x,y,q)=\sum_{n=0}^\infty\frac{(-qx)^nq^{\binom n2}}{(q;q)_n(qy;q)_n}\;,
\end{equation}
with the $q$-product notation $(t;q)_n=\prod\limits_{k=0}^{n-1}(1-tq^k)$. The function $H(x,y,q)$ is therefore given in terms of the $q$-Bessel function ${}_1\phi_{1}$ as $H(x,y,q)={}_1\phi_{1}(0;qy;q,qx)$. Therefore we have the explicit exact solution
\begin{equation}
F(c,x,y,q)=\frac1{1-c\left[
x+y
\frac{\displaystyle\sum\limits_{n=0}^\infty\frac{(-q^2x)^nq^{\binom n2}}{(q;q)_n(qy;q)_n}}{\displaystyle\sum\limits_{n=0}^\infty\frac{(-qx)^nq^{\binom n2}}{(q;q)_n(qy;q)_n}}
\right]}\;.
\end{equation}
The solution simplifies considerably in the case of unweighted area, i.e. $q=1$. Here, the functional equation (\ref{functeq}) reduces to a quadratic equation for $S(x,y,1)$, and it follows that
\begin{equation}
\label{q=1}
F(c,x,y,1)=\frac1{1-\frac c2\left[x+y+1-\sqrt{x^2-2xy+y^2-2x-2y+1}\right]}\;.
\end{equation}
It is also useful in calculations to write the generating function in terms of a continued fraction. Solving the functional equation (\ref{functeq}) for $S(x,y,q)$, a continued fraction expansion becomes apparent, and we find
\begin{equation}
F(c,x,y,q)=\cfrac1{1-cx-\cfrac{cy}{1+y-qx-\cfrac{y}{1+y-q^2x-\cfrac{y}{1+y-q^3x-\cfrac{y}{1+y-q^4x-\ldots}}}}}\;.
\end{equation}

\section{Phase Diagram}

\begin{figure}[ht]
\includegraphics[height=8cm]{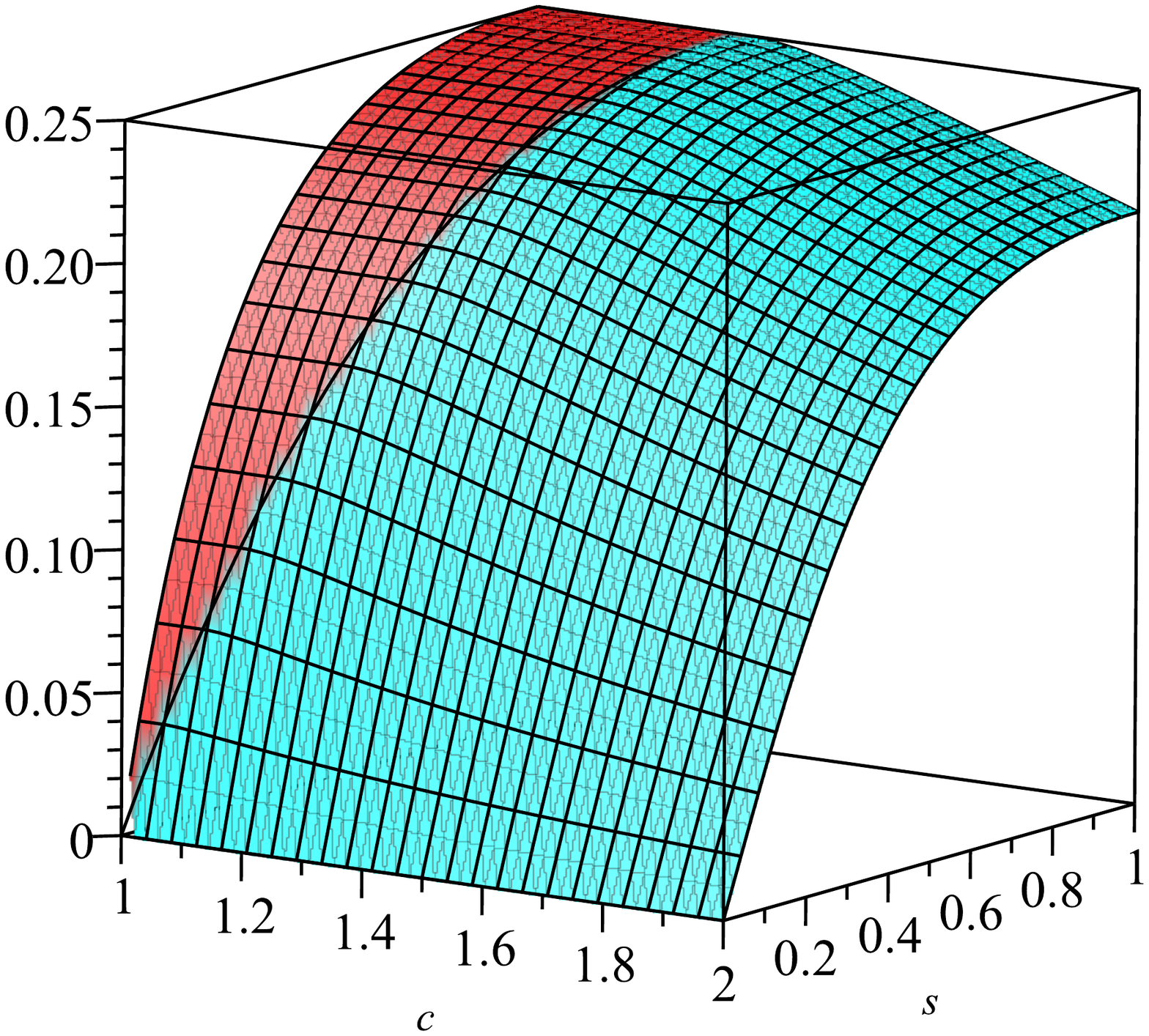}
\caption{A plot of the singularity closest to the origin of the generating function with $q=1$. The red (darker) part of the surface arises from the square root singularity $t_r$ and is associated with the unbound phase. The cyan (lighter) part of the surface iarises from  the simple pole $t_p$ and is associated with the bound phase. The black line indicates the phase transition.}
\label{fig2}
\end{figure}

We begin our analysis with considering the case $q=1$. From Equation (\ref{q=1}) we find that the generating function $G(c,x,y,1)$ has a square root and a simple pole. In terms of the original variables $c$, $t$ and $s$ this implies that we find a square root singularity at
\begin{equation}
\label{tr}
t_r=\frac s{(1+s)^2}
\end{equation} 
independent of the value of $c$. There is also a simple pole which we obtain by setting the denominator of (\ref{q=1}) to zero, from whence we find
\begin{equation}
t_p=\frac{(1-c)(1+s^2)+\sqrt{[(1+s^2)^2c-(1-s^2)^2](c-1)}}{2cs}\;.
\label{tp}
\end{equation}
This expression is found from solving $1/G(c,t/s,ts,1)=0$, which can be expressed as
\begin{equation}
{\sqrt { \left(  \left( s-1 \right) ^{2}t-s \right)  \left( 
 \left( s+1 \right) ^{2}t-s \right) }}=ts^2+s+t-\frac{2s}c\;.
\end{equation}
Note that both sides of this equation need to be non-negative, in particular ,
\begin{equation}
c\geq\frac{2s}{(s^2+1)t+s}\;.
\end{equation}
Also note that $0\leq t\leq t_r$ 
which implies that Equation (\ref{tp}) is only valid if
\begin{equation}
c\geq\frac{(s+1)^2}{s^2+s+1}\equiv c_s\;.
\label{cs}
\end{equation}
%
%
%
In fact, when $c=c_s$ then $t_p=t_r$. Putting this together, we find
\begin{equation}
t_c(c,s,1)=\begin{cases} 
 \dfrac s{(1+s)^2}\;,& c\leq \dfrac{(s+1)^2}{s^2+s+1}\;,\\ \dfrac{(1-c)(1+s^2)+\sqrt{[(1+s^2)^2c-(1-s^2)^2](c-1)}}{2cs}\;,& c\geq \dfrac{(s+1)^2}{s^2+s+1} \;.
\end{cases}
\end{equation}
The behaviour of $t_c(c,s,1)$ is illustrated in Figure \ref{fig2}. 

\begin{figure}[ht]
\hspace*{-1cm}\includegraphics[height=6cm]{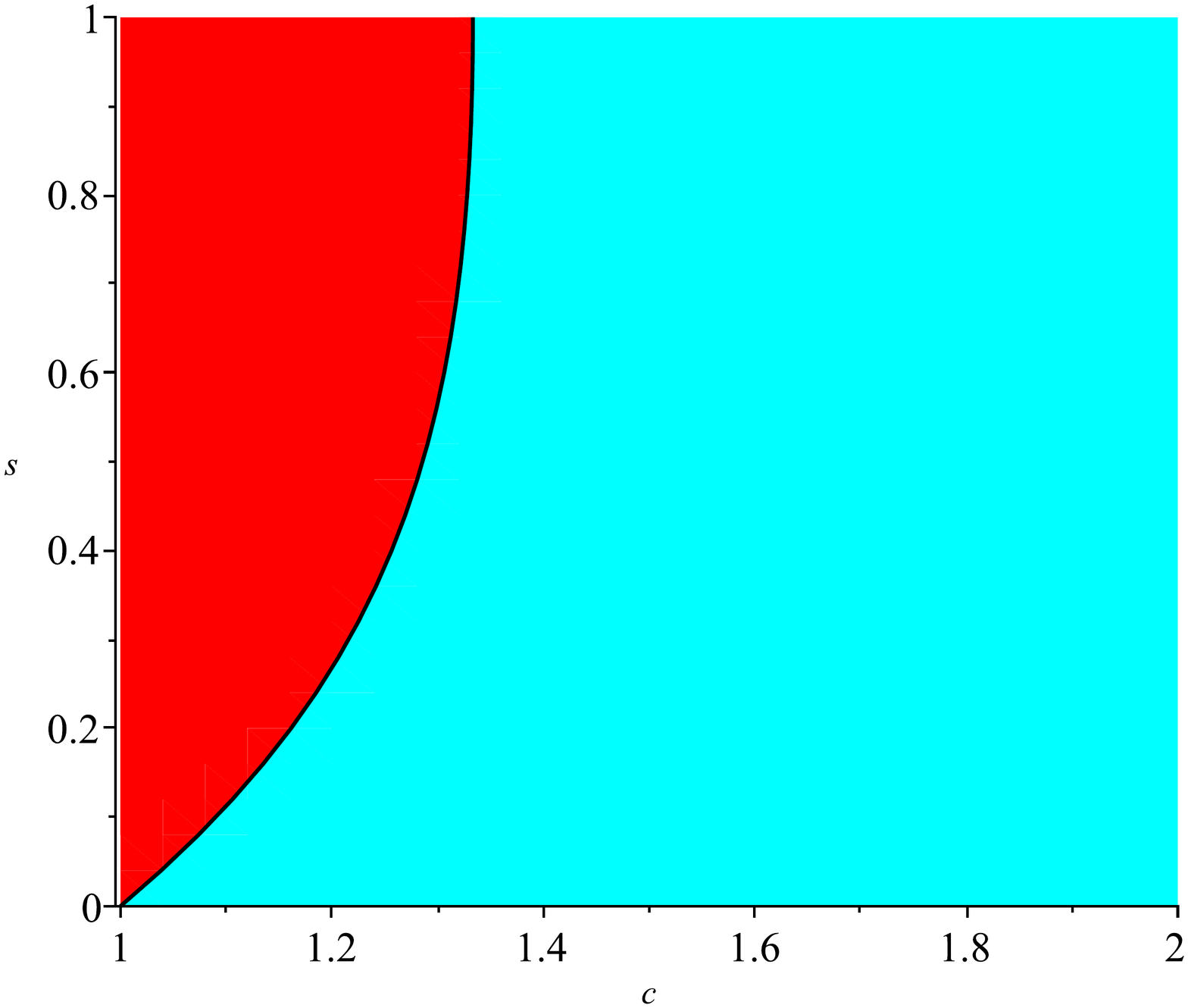}\includegraphics[height=6cm]{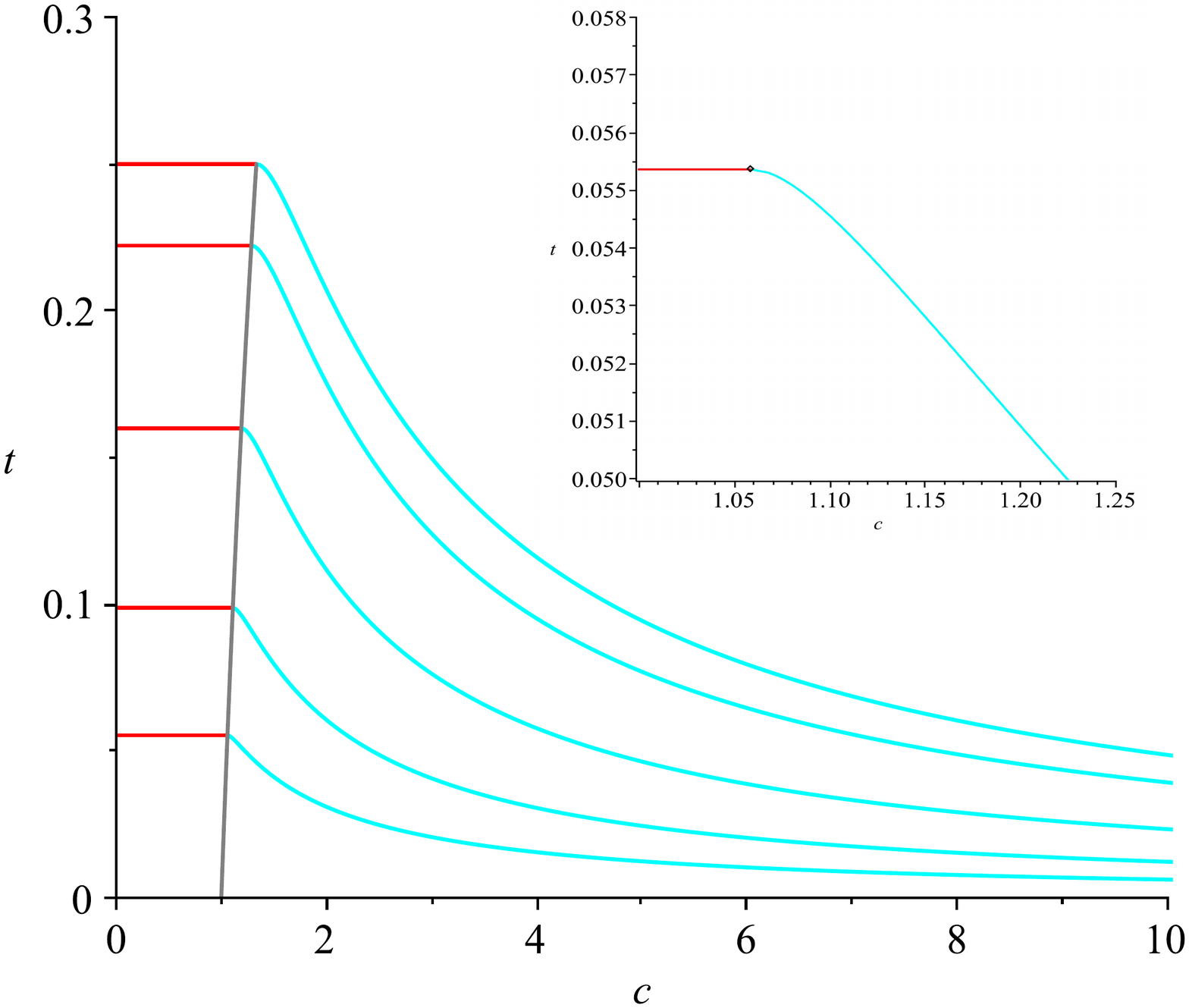}
\caption{On the left hand side is the phase diagram of the model when $q=1$. The red area indicates the unbound (darker) phase while the cyan (lighter) area indicates the bound phase. On the righthand side the closest singularity to the origin of the generating function is plotted against $c$ for various values of the fugacity associated with the force for $s=1,1/2,1/4,1/8,1/16$, from top to bottom, with an inset for $s=1/16$. The inset demonstrates that despite the transition looking sharper for increasing $s$ it is still smooth (not first order).}
\label{fig3}
\end{figure}

We next consider the density of contacts  
\begin{equation}
{\cal M} (c,s,q) \equiv \lim\limits_{n\to\infty}\dfrac1n\langle m_c\rangle_n
=-\frac{\partial\log t_c(c,s,q)}{\partial\log c}\;.
\end{equation}
We find
\begin{equation}
{\cal M}(c,s,1) 
=\begin{cases} 
0\;,& c\leq \dfrac{(s+1)^2}{s^2+s+1}\;,\\ \dfrac{c-2}{2(c-1)}+\dfrac{c(1+s^2)}{2\sqrt{[(1+s^2)^2c-(1-s^2)^2](c-1)}}\;,& c\geq \dfrac{(s+1)^2}{s^2+s+1} \;.
\end{cases}
\end{equation}
One can therefore understand this transition as the binding-unbinding transition of the two walks, where for small values of $c$ the two walks remain unbound with zero density of contacts, while for large values of $c$ there is a finite density of contacts. At the phase transition, the density of contacts starts to increase linearly:
\begin{equation}
{\cal M} \sim\frac2{c_s^3}\frac{(1+s)^2}s  (c-c_s)
\label{Mccs}
\end{equation}
for $c>c_s$.
Putting all this together leads to the phase diagram on the left hand side of Figure~\ref{fig3}. 
On the right hand side of Figure~\ref{fig3} are plots of the closest singularity to the origin for various fixed values of $s$ demonstrating the behavior as $s$ is varied: while the transition looks sharper at small values of $s$ closer inspection reveals that the transition is still smooth albeit over a smaller range of $c$.

\begin{figure}[ht]
\includegraphics[height=8cm]{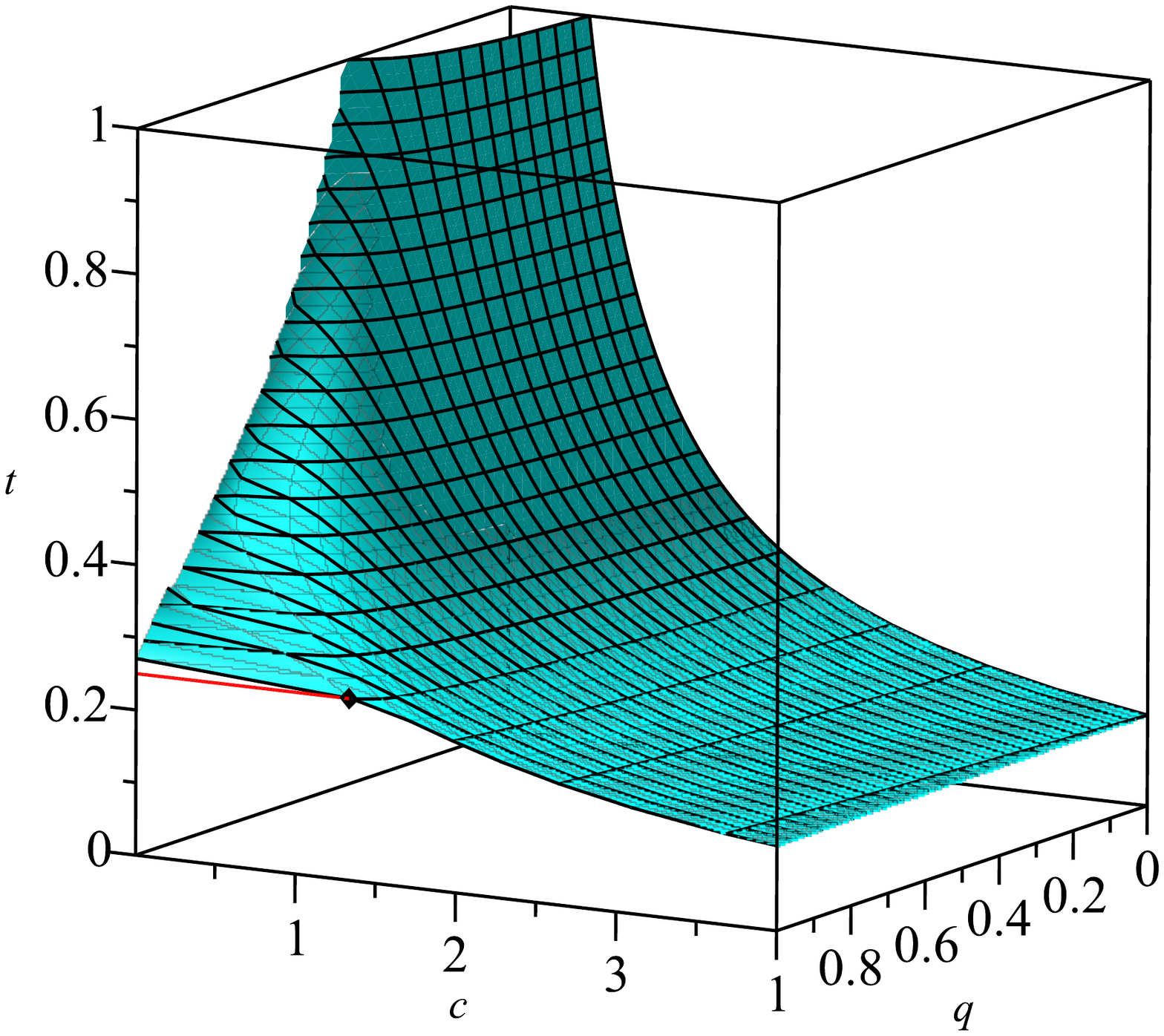}
\caption{ A plot of the singularity closest to the origin of the generating function with $s=1$ for varying $q$ and $c$.}
\label{fig4}
\end{figure}

\begin{figure}[ht]
\includegraphics[height=6cm]{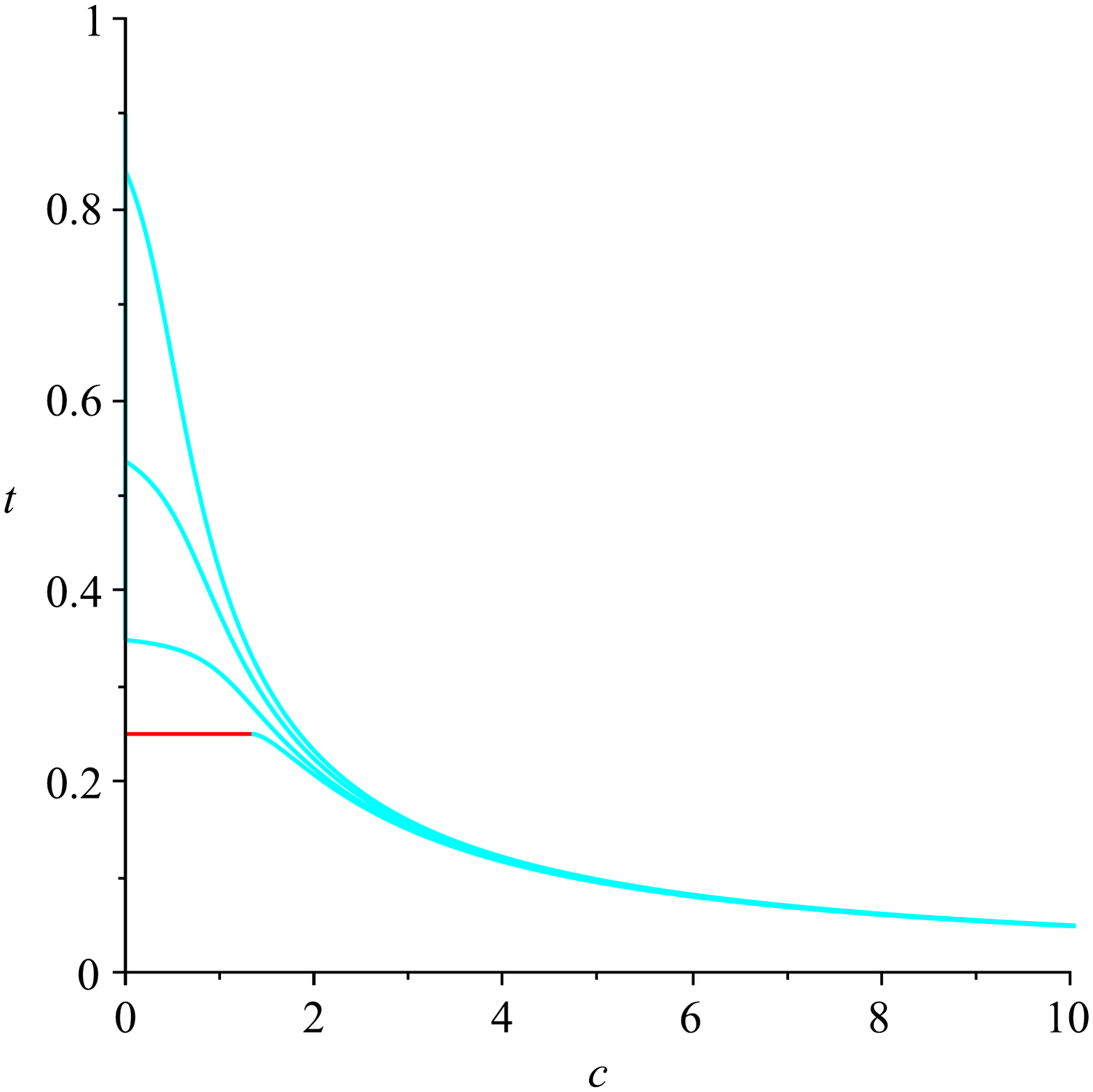}\includegraphics[height=6cm]{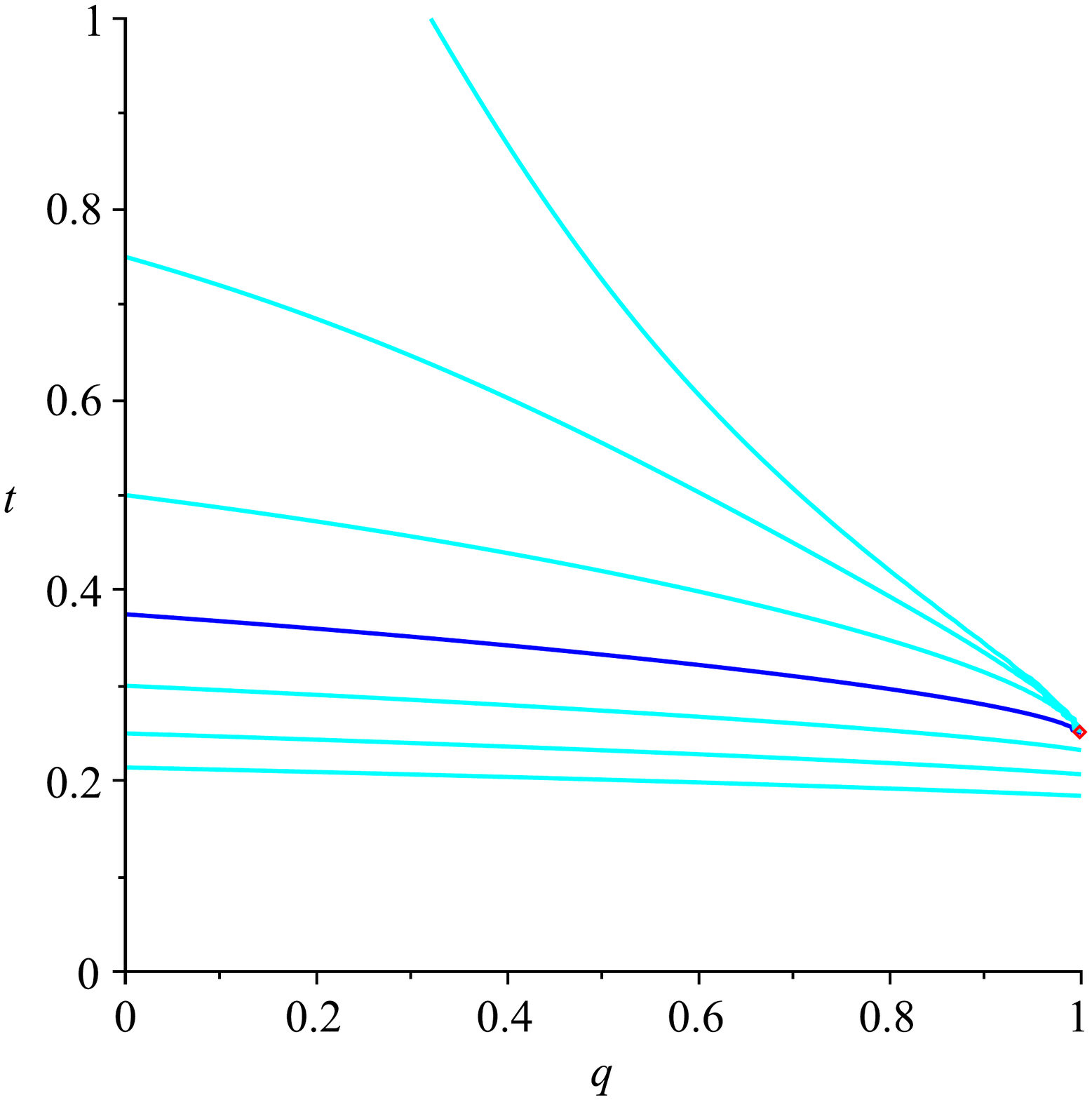}
\caption{On the left hand side the closest singularity to the origin of the generating function is plotted against $c$ for various values of the fugacity associated with the pressure for $q=1,0.9,0.7,0.5$ from bottom to top. On the right hand side the closest singularity to the origin of the generating function is plotted against $q$ for various values of the fugacity associated with the binding for $c=1/3,2/3,1,4/3,5/3,2, 7/3$ from top to bottom. Note that for $c=1/3,2/3,1,4/3$ that as $q\rightarrow 1$ the value of $t_c$ approaches the common value of $1/4$. We have highlighted the curve $t_p(4/3,1,q)$ associated with $c=4/3$ in dark blue as $4/3$ is the value of the binding fugacity at $q=1$.}
\label{fig5}
\end{figure}

This concludes our discussion of the $q=1$ case. We now consider the effect of including area weights. When $q>1$ configurations with large area dominate. These configurations have an area that grows quadratically in length, so that $t_c$ becomes zero and the bound phase disappears. Conversely, when $q<1$, we find for any values of $s$ and $c$ a positive density of contacts, and the unbound phase disappears.
The dominating singularity in the generating function is always a simple pole. The location $t_p(c,s,q)$ of this pole is easiest given implicitly as a continued fraction expansion of $c$ in terms of $t_p$, $s$, and $q$.
\begin{equation}
\label{cfrac}
c=\cfrac1{\dfrac{t_p}s+\cfrac{{t_p}s}{1+{t_p}s-q\dfrac {t_p}s-\cfrac{{t_p}s}{1+{t_p}s-q^2\dfrac{t_p}s-\cfrac{{t_p}s}{1+{t_p}s-q^3\dfrac {t_p}s-\cfrac{{t_p}s}{1+{t_p}s-q^4\dfrac {t_p}s-\ldots}}}}}\;.
\end{equation}
This expansion can be used for numerical evaluations. In Figures~\ref{fig4} and \ref{fig5} we plot various views of the singularity closest to the origin of the generating function highlighting different features of the behaviour using equation~(\ref{cfrac}). 
In Figure~\ref{fig4} we have plotted the singularity closest to the origin of the generating function with $s=1$ for varying $q$ and $c$. On the left hand side of Figure~\ref{fig5} we have plotted the singularity closest to the origin of the generating function with $s=1$ for various values of the fugacity associated with the pressure against the binding fugacity $c$: note that the singularity moves away from the origin as $q$ is lowered for all $c$. Moreover, for all $q<1$ it is a smooth function of $c$ highlighting the loss of any fixed $q$ phase transition. On the righthand side of  Figure~\ref{fig5} we have plotted the singularity closest to the origin of the generating function with $s=1$ for various values of the fugacity associated with the binding against the pressure fugacity $q$. Notice that for all $c$ the singularity moves away from the origin on decreasing $q$. For $c\leq4/3=c_s$ the curves approach the $q=1$ value of the singularity $t_c=1/4$ in a cusp-like fashion. In the scaling analysis section we calculate the associate scaling form for the critical value of $c=4/3$: the curve is highlighted in the figure. For $c>4/3$ the curves approach the $q=1$ value smoothly.


\section{Scaling Analysis}
For simplicity in this section we restrict ourselves to $s=1$ but all the analysis is similar for $s \neq 1$ noting the change of the binding transition from $c=4/3$ for $s=1$ to that given by $c_s$ as in equation~(\ref{cs}).

As $t\to1/4$ and $q=1-\epsilon\to1^{-}$ such that $(1-4t)\epsilon^{-2/3}$ is constant, an asymptotic expansion for the generating function of staircase polygons has been derived \cite{prellberg1995d-a}.
Using
\begin{equation}
S\sim\frac14+4^{-2/3}\epsilon^{1/3}\frac{\Ai'(4^{1/3} (1-4t)\epsilon^{-2/3})}{\Ai(4^{1/3}(1-4t)\epsilon^{-2/3})}
\end{equation}
we find that asymptotically the singularity $t_p(c,1,q)$ of the generating function $G(c,1,q,t)$ closest to the origin is given implicitly  by
\begin{equation}
c\sim\frac1{\dfrac34+4^{-2/3}\epsilon^{1/3}\dfrac{\Ai'(4^{1/3} (1-4t_p)\epsilon^{-2/3})}{\Ai(4^{1/3}(1-4t_p)\epsilon^{-2/3})}}\;;
\end{equation}
this is the asymptotic form of Eqn. (\ref{cfrac}). From this asymptotic form one can deduce that at $c=4/3$ the location of the pole $t_p(4/3,1,q)$ as shown in dark blue in Figure~\ref{fig5}, and hence the free energy, behaves as
\begin{equation}
t_p(4/3,1,q) \sim \frac{1}{4} - a'_1 4^{-4/3}(1-q)^{2/3}
\end{equation}
where $a'_1=-1.0187...$ is the location of the first zero of the function $\Ai'(z)$. This implies that at this special value of the binding fugacity the inflation transition is also critical with an average area that diverges when $q\rightarrow 1^-$ as
\begin{equation}
{\cal A}(4/3,1,q) = \lim_{n\to\infty}\dfrac1n\langle a \rangle_n \sim -a'_1\frac{2^{1/3}}3(1-q)^{-1/3}
\label{Aqcs}
\end{equation}
where we have defined 
\begin{equation}
{\cal A} (c,s,q) \equiv \lim\limits_{n\to\infty}\dfrac1n\langle a\rangle_n
=-\frac{\partial\log t_c(c,s,q)}{\partial\log q}\;.
\end{equation}
Similarly, the density of contacts tends to zero as
\begin{equation}
{\cal M}(4/3,1,q)  \sim -\frac1{a'_1}\frac3{4^{2/3}}(1-q)^{1/3}\;.
\label{Mqcs}
\end{equation}
For $c>4/3$ 
\begin{equation}
 {\cal A}(c,1,1) =  \frac{2(c-1) +\sqrt{c(c-1)}}{3c-4}
 \end{equation}  remains bounded with
\begin{equation}
{\cal A}(c,1,1)  \sim \frac{4}{9} (c-4/3)^{-1}
\label{Accs}
\end{equation}
as $c$ approaches $4/3^+$. For $c<4/3$  in the unbound phase $ {\cal A}(c,1,1)$ is undefined.

It is also of interest to calculate the related scaling forms  fixed at different values of $c$ and $q$ in each of the phases as shown in the schematic of the phase diagram (Figure~\ref{phase-diagram}): 
\begin{figure}[ht!]
\includegraphics[height=8cm]{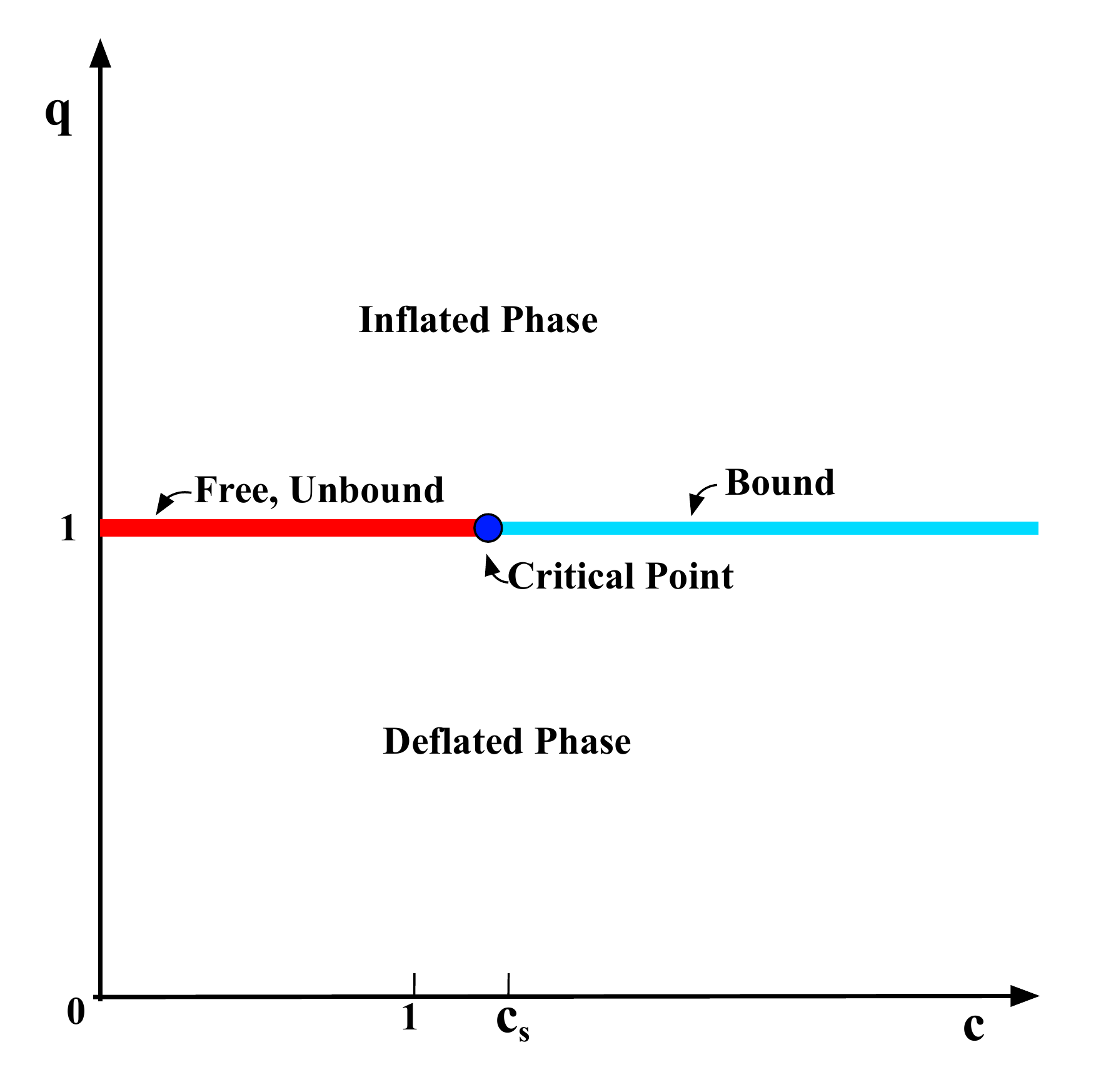}
\caption{A schematic of the phase diagram with each of the phases labelled, for our model at fixed but arbitrary $s$: the only change is the value of $c_s$ as given by equation~(\ref{cs}).}
\label{phase-diagram}
\end{figure}
the scaling forms appear in Table~\ref{exptable}. These forms can be deduced from our results and previously published ones. For $q=1$ the behaviour of the average number of contacts can be seen easily from our results and is well known. For the bound and deflated phases the existence of the non-zero limits of ${\cal A}$ and ${\cal M}$ implies the $n^1$ scaling. Previous work \cite{richard2002a-a} on area moments of staircase polygons gives the unbound phase scaling of the area as $n^{3/2}$. We have extended this result to $c=4/3$. The inflated phase can be easily deduced from previous work\cite{prellberg1999a-:a}.
\begin{table}[ht]
\begin{center}
\begin{tabular}{ |c | c | c | c | } 
\hline
Phase region & Phase & $\langle m \rangle_n$ & $\langle a \rangle_n$\\
\hline
$q< 1$ & Deflated  & $n^1$ & $n^1$\\ 
\hline 
$q> 1$ & Inflated  & $n^0$ & $n^2$\\ 
\hline 
$q=1$ $c<c_s$& Free, unbound  & $n^0$ & $n^{3/2}$\\ 
\hline 
$q=1$ $c=c_s$ & Critical  & $n^{1/2}$ & $n^{3/2}$\\ 
\hline 
$q=1$ $c> c_s$ & Bound  & $n^1$ & $n^1$\\ 
\hline 
\end{tabular} 
\end{center}
\caption{Scaling of the mean number of contacts and number of area plaquettes at fixed $c$ and $q$ in each phase of the phase diagram.}
\label{exptable}
\end{table}

To conclude let us notice that the exponents relating the behaviours in equations~ (\ref{Aqcs}), (\ref{Accs}), (\ref{Mqcs}), and (\ref{Mccs}) of both ${\cal A}(c,1,q)$ and ${\cal M}(c,1,q)$ on approaching $(c,q)\rightarrow (4/3,1)$ differ by a crossover exponent of $1/3$. It would be interesting in future work to see whether any quantities can be written in terms of the scaling combination $(c-4/3) (1-q)^{- 1/3}$. Note the binding value $4/3$ would be more generally replaced by $c_s$.

\section*{Acknowledgements}

Financial support from the Australian Research
Council via its Discovery Projects scheme (DP160103562)
is gratefully acknowledged by one of the authors, A L Owczarek, who also 
thanks the School of Mathematical Sciences, Queen Mary University of London 
for hospitality. 


\end{document}